# Effect of UHV annealing on morphology and roughness of sputtered Si(111)-(7×7) surfaces

Jagadish Chandra Mahato [a, *], Anupam Roy [b, *], Rajib Batabyal [c], Debolina Das [d], Rahul Gorain [b], Tuya Dey [a], and B. N. Dev [e]

[a]Department of Physics, Siksha Bhavana, Visva-Bharati, Santiniketan, Bolpur (W.B.) 731235, India
[b]Department of Physics, Birla Institute of Technology Mesra, Ranchi (Jharkhand) 835215, India
[c]UGC-DAE Consortium for Scientific Research, University Campus, Khandwa Road, Indore (M.P.) 452001, India
[d]Department of Physics, Haldia Government College, Haldia, Debhog, Purba Medinipur (W.B.) 721657, India
[e]Centre for Quantum Engineering, Research and Education (CQuERE), TCG Centres for Research and Education in Science and Technology (TCG CREST), Sector V, Salt Lake, Kolkata 700091, India

## ABSTRACT

Ar[+] ion has been used regularly for the cleaning of semiconductor, metal surfaces for epitaxial nanostructures growth. We have investigated the effect of low-energy Ar[+] ion sputtering and subsequent annealing on the Si(111)-(7×7) surfaces under ultrahigh vacuum (UHV) condition. Using *in-situ* scanning tunnelling microscopy (STM) we have compared the morphological changes to the Si(111)-(7×7) surfaces before and after the sputtering process. Following 500 eV Ar[+] ion sputtering, the atomically flat Si(111)-(7×7) surface becomes amorphous. The average root mean square (rms) surface roughness ($\sigma_{avg}$) of the sputtered surface and that following post-annealing at different temperatures (500 – 700°C) under UHV have been measured as a function of STM scan size. While, annealing at ~500°C shows no detectable changes in the surface morphology, recrystallization process starts at ~600°C. For the sputtered samples annealed at temperatures $\geq$ 600°C, $log\sigma_{avg}$ varies linearly at lower length scales and approaches a saturation value of ~0.6 nm for the higher length scales confirming the self-affine fractal nature. The correlation length increases with annealing temperature indicating gradual improvement in crystallinity. For the present experimental conditions, 650°C is the optimal annealing temperature for recrystallization. The results offer a method to engineer the crystallinity of sputtered surface during nanofabrication process.

[*]Corresponding author.
Email addresses: jagadishc.mahato@visva-bharati.ac.in (J. C. Mahato), royanupam@bitmesra.ac.in (A. Roy)

## 1. Introduction

Ion-beam-assisted modification of semiconductor surfaces has received enhanced interest in recent years, [1-4] due to the potential of this bottom-up procedure for applications in Nanotechnology. $Ar^+$ ion has been used regularly for the cleaning of semiconductor, metal surfaces for growth of epitaxial nanostructures. Ion bombardment has also been utilized for smoothening and roughening of the surfaces that modifies the effective surface free energy, thereby altering the morphology of the thin film grown on these modified substrates [5-10]. Several studies have focused on studying the parameters, *e.g.*, energy, ions, fluence, implantation angle, *etc.*, that govern the alterations of the target substrate [7-9]. Besides its diverse applications in materials science, *e.g.*, implantation, doping, etching, surface preparation, *etc.*, [1-4, 11-18] ion-beam sputtering gives rise to new features in the wider context of pattern formation at sub-micrometer scales [14, 19-23]. Ions with lower energies (with energies in keV or below) can be used to produce self-organized patterns on surfaces such as dots, ripples, pits, *etc.*, at nanoscale [10, 24-28]. Fabrication of nanodots using low-energy ion irradiation holds promising applications because of the possibility of the generation of large-area well-arranged arrays of dots in a single technological step [12]. These nanoscale patterns have found several applications in optoelectronic and photonic devices, suitably reconstructed substrates for scanning tunneling microscopy (STM) studies, good oxidation surface, and strain-free structured templates, *etc.* [5, 6, 29]. Along with the experimental investigations, there have been attempts to develop the theoretical understanding of nano-patterning dynamics [26, 30-33]. Interestingly, pattern formation process seems to be largely independent of the type of ions [4, 34].

Previously, Seoo *et al.* have used ion sputtering to generate dot patterns in Ni separated about 250-300 nm that were used to separate DNA molecules [35]. Facsko *et al.* has reported formation of nanodots using ion beam sputtering that was extended to the quantum dot array fabrication [36, 37]. However, the crystallinity of the target substrate following the sputtering process is a concern since this process can induce considerable amorphization on the surface. Therefore, an improvement in the crystallinity of the surface following the sputtering process is necessary for this process to be fully applicable. Nano-pattern formation on different crystallographic silicon surfaces using $Ar^+$ ion sputtering with different energies have been investigated previously [38-42]. Zandvliet *et al.* reported the effect of Ar[+] ion beam (3 keV) impact on the Si(111) surfaces [43], at room temperature (RT). Ghose *et al.* investigated the surface and subsurface defect formations *via* low energy (200 eV) Ar[+] ion irradiation on Si(111)-(7×7) surface [13]. With increasing ion energy, sputtering of silicon leads to a greater number of broken bonds, producing larger collision cascades and thus more complexity in ejection pathways [44]. Brown et. al. have investigated the spontaneous regular patterned morphological evolution of the Si(111) surface during energy 250–1200 eV, Ar[+] ion beam etching at elevated temperature 500–750 °C oblique incidence 60° from normal, and the influence on morphological evolution by ion fluence and ion flux [45]. Previously Ar[+] ion implantation induced disorder and recovery through annealing have been studied *via* Rutherford Backscattering Spectroscopy (RBS) or Raman spectroscopy [46, 47]. However, a more detailed study to control the sputtering-induced damage reduction in silicon substrates under ultrahigh vacuum (UHV) condition is still necessary. In order to investigate the atomistic processes on the ion-beam-irradiated surface at atomic scale, in-situ experimental arrangement is inevitable. The investigation of surface roughness parameters is also very important, as they influence the growth behaviour of nanostructures on such surfaces [7, 9]. There are report in the past on structures and energies of different reconstructions in Si(111) surface, using the density functional theory, which reveals (7×7) surface possesses the lowest surface free energy compared to the surfaces reconstructed in other forms [48].

Although extensively investigated, the complexity in (7×7) reconstruction of Si(111) surface attracts attention to study different fundamental properties of surface science as well as for promising novel applications [49-51]. Recently, Hu *et al.*, using artificial neural-network potential carried out large-scale simulations which suggests a critical collective vacancy diffusion process leading to a sequence of selective dimer, corner-hole, stacking-fault, and dimer-line pattern formation, making completion of the 7×7 reconstruction [49]. Guo *et al.*, reported that Si(111)-(7×7) offers a facile strategy to obtain and utilize identical metal clusters such as $Pd_6$ cluster for practical clean energy production which selectively converting $CO_2$ to high-value $C_2$–$C_4$ hydrocarbons and alcohols [51]. In the current work, we investigate the morphological changes of Si(111)-(7×7) surface due to Ar[+] ion sputtering *via* in situ STM experiments under UHV condition. The detailed investigation regarding the roughness exponent and correlation length as a function of temperature is the main focus of this study, since the self-affine fractal morphology of the surface plays a pivotal role in further growth of thin films or quantum nanostructures on top of it. The physical and chemical properties of the material depends crucially on the fractal nature of the underlying substrate [52]. The underlying surface morphology can influence the growth of nanostructures, thereby, controlling the optical [53], mechanical [54], electrical [55], magnetic [56], and many other properties. A detailed investigation of the roughness parameters and fractal characteristics can offer an effective way of engineering the crystallinity of the sputtered surface so as to suit the criteria of any specific device application.

We have used atomically clean Si(111)-(7×7) surfaces prepared under the UHV and irradiated by 500 eV Ar$^+$ ions. The amorphization of the surface following the sputtering process has been confirmed using *in situ* STM. To study the recrystallization of the sputtered surface, we employ post-annealing treatment *in situ* under UHV. Analyzing the STM images, we present a comparison of the morphologies and surface roughness parameters, such as the average root-mean-square (rms) surface roughness ($\sigma_{avg}$), roughness exponent, correlation length *etc.*, of the sputtered surface following annealing at different temperatures (500 °C– 700 °C) under UHV condition.

## 2. Experimental details

The experiments were performed in a custom-made molecular beam epitaxy (MBE) growth chamber under UHV condition (base pressure ~ 5.2×10$^{-11}$ mbar). Post growth investigations of the samples were carried out by *in situ* STM at RT. The base pressure in the STM chamber was 2.3×10$^{-10}$ mbar. Substrates used in the experiments were P-doped, n-type Si(111) wafer with resistivity of 10-20 Ω-cm. Atomically clean, reconstructed Si(111)-(7×7) surfaces were prepared by the standard heating and flashing procedure [57]. The samples were first degassed at ~700 °C for about 14 hours, and then flashing the substrate at ~1250 °C for one minute. The samples were then quickly brought down to ~780 °C and annealed at this temperature for 30 minutes. Finally, the samples were allowed to cool down slowly to RT.

Si(111)-(7×7) surface reconstruction was observed using *in-situ* STM. All of the STM images were captured with a bias voltage, $V_b = 1.9$ V and a tunneling current, $I_t = 0.2$ nA.

The sample was then transferred back to the MBE chamber for *in situ* sputtering with 500 eV Ar$^+$ ions of fluence $1.1 \times 10^{16}$ ions/cm$^2$, incident at an oblique angle of 45° with the Si(111) surface normal. High purity Argon gas was fed through a nozzle valve in a controlled manner. Argon ions are created in the ionizer before the ions enter the vacuum chamber. Argon ions are accelerated by the electric field applied at the sample stage. During the sputtering process, the pressure in the MBE chamber went up to ~$1.0 \times 10^{-5}$ mbar due to insertion of Ar gas. Following the Ar$^+$ ion sputtering of the Si(111)-(7×7) surface, the sample was characterized again *via* STM. Subsequently, the sample underwent in situ thermal annealing under UHV (in the MBE chamber) at different temperatures: ~500 °C, ~600 °C, ~650 °C and ~700 °C for 1 hour. For the annealing at a specific temperature, the sample temperature was raised from RT to the desired temperature at a ramp of 15 °C/min using resistive heating up to ~ 500°C and beyond at a ramp of ~25 °C/min using the direct heating current. After the completion of the annealing process, the sample was cooled down to RT at a rate of ~25 °C/min. The pressure during annealing was below $1.0 \times 10^{-10}$ mbar. The samples were characterized via STM at RT after each annealing process.

## 3. Result and discussion

### 3.1. Effect of Ar$^+$ ion sputtering on clean Si(111)-(7×7) surfaces

Figure 1(a, b, and c) show STM images of a clean Si(111)-(7×7) surface of scan areas 1000×1000 nm$^2$, 200×200 nm$^2$, and 50×50 nm$^2$, respectively. Any improvement in crystallinity can easily be realized from the high resolution STM image at atomic scale. High resolution STM image of scan area 50×50 nm$^2$ [Figure 1(c)] provides the evidence of the flat surface areas with the periodicity and regular, repetitive atomic arrangement. An atomically resolved (7×7) reconstructed unit cell with 12 adatoms of the faulted and unfaulted halves is shown in the (bottom right) inset of Figure 1(c). Figure 1(d, e, and f) show corresponding STM images of the same Si(111)-(7×7) surface following 500 eV Ar$^+$ ion sputtering. From STM studies, it is clear

that following the sputtering process, an atomically clean single-crystal surface of Si(111)-(7×7) has turned into an amorphous surface. High resolution STM investigation reveals that there is no atomic periodicity on the sputtered surface [Figure 1(f)]. Height profiles across the steps [as marked in Figure 1(a) and (d)] are shown in Figure 1(g) and (h), respectively. Fourier transforms (FTs) corresponding to Figure 1(c) and (f) are shown in Figure 1(i) and (j) respectively. The disappearance of (7×7) ordered spots is also evident from the FTs patterns.

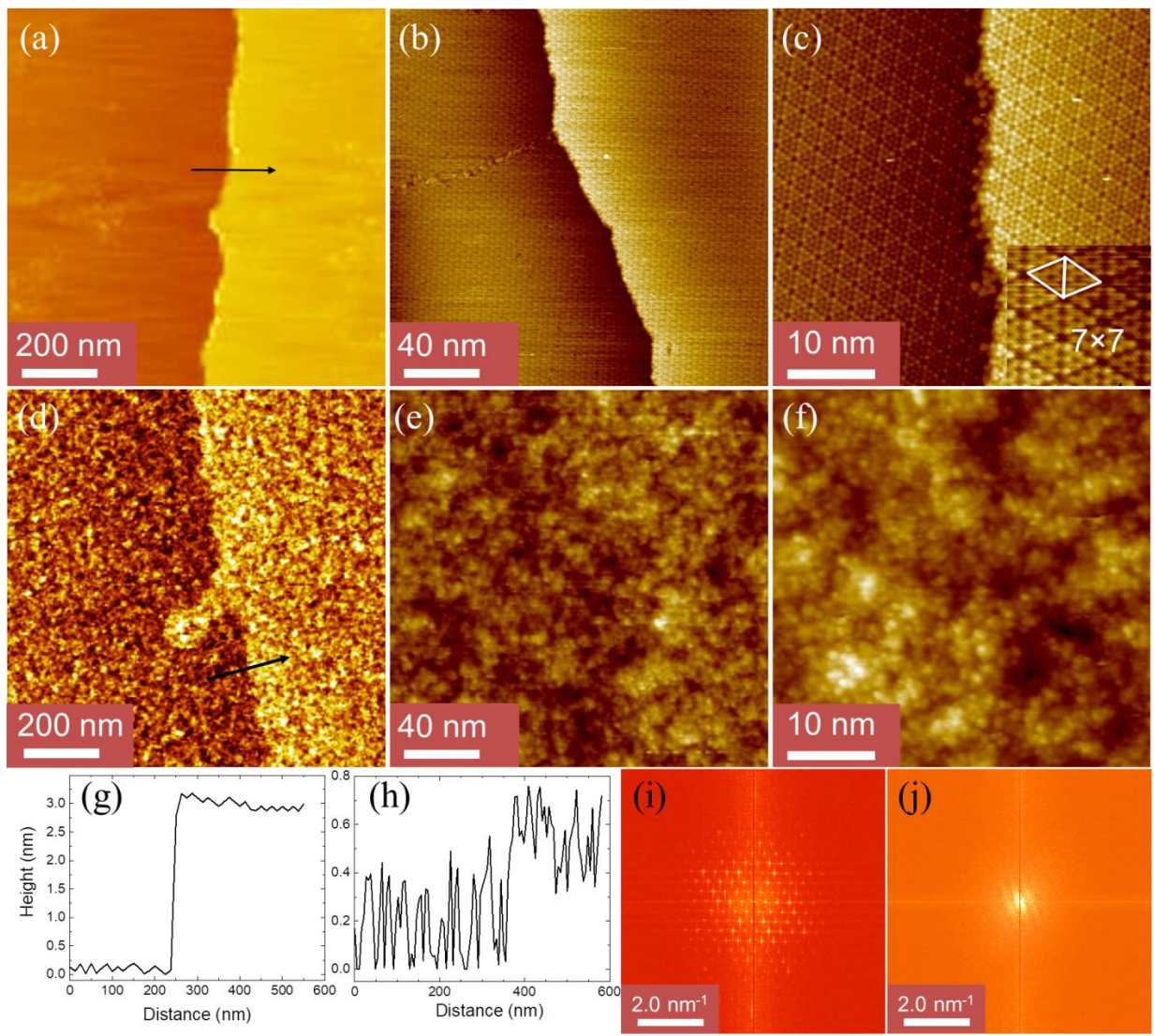

**Figure 1**: STM images of atomically clean Si(111)-(7×7) surfaces of scan areas (a) 1000×1000 nm², (b) 200×200 nm², and (c) 50×50 nm² are presented. The atomic resolution image of (7×7)

structure is shown as inset in (c) (bottom right). Corresponding STM images of the surface after $Ar^+$ ion sputtering are shown in (d, e, f). Figures (g) and (h) show the line profiles captured across the step edge in (a) and (d), respectively. Fourier transforms (FTs) corresponding to figures (c) and (f) are shown in figures (i) and (j), respectively. Diffraction spots corresponding to periodic (7×7) structure disappear completely following the sputtering process.

*3.2. Recrystallization of amorphous Si(111) surfaces via annealing*

Figure 2 shows STM images, of scan areas 1000×1000 $nm^2$, 200×200 $nm^2$ and 50×50 $nm^2$, respectively, from sputtered-Si(111) surfaces following annealing at different temperatures. More STM images captured systematically from larger scan areas to smaller scan areas have been shown in the supplementary material in Figure S1 and S2 for the samples annealed at 650 °C and 700 °C. STM studies reveal that the sputtered surface [in Figure 1(d), (e), and (f)] remains amorphous after annealing at ~500 °C *via* resistive heating [Figure 2(a), (b) and (c)] with no detectable changes in the surface morphology. Figure 2 (e, f, g); (i, j, k) and (m, n, o) show STM images of the sputtered-Si(111) surfaces annealed at ~600 °C, ~650 °C and ~700 °C, with the scan areas of 1000×1000 $nm^2$ [Figure 2 (e, i, m)], 200×200 $nm^2$ [Figure 2 (f, j, n)], and 50×50 $nm^2$ [Figure 2 (g, k, o)]. Line profiles recorded on the STM images for the samples annealed at 500-700 °C have been shown in Figure S3 supplementary material. Upon annealing at ~600 °C under UHV the amorphous surface starts recrystallizing, as evident from the STM investigations [shown in Figure 2 (e) (f), and (g)]. This has also been observed by Yang et al., where they observed the transition from Ar-sputtered unreconstructed surface into the (7×7) surface following an annealing at higher temperature [58]. High resolution STM image in Figure 2(g), illustrates many flat-plateau of Si(111) surfaces with top surface layer atoms reconstructed as (7×7) units similar to that of a clean Si(111) surface. The average size of these plateaus is a few hundred $nm^2$, as estimated from STM. At different locations, especially at the edges of these plateaus, there is presence of silicon nanodots or nanoislands. The scanning tunneling spectroscopy (STS) performed on one of these nanodots is shown in Figure S4 in supplementary material. Thus, upon annealing at ~600 °C, the sputtered-Si(111) surface undergoes a transition from amorphous to crystalline. Although the surface reconstruction is predominantly of (7×7) structure, presence of (2×2) reconstructed chain like structure and (5×5) reconstructed surface structure at some locations is also observed. Similarly, STM investigations, carried out on the samples annealed at ~650 °C and ~700 °C, reveal that the average plateau size increases with increasing annealing temperatures [as shown in Figures 2 (i, j, k) and (m, n, o)]. The average size of the nanodots/nanoislands also increases upon annealing at 650°C and 700°C. These nanodots/nanoislands are possibly formed by the agglomeration of randomly distributed silicon atoms produced due to the sputtering process. The terrace edges of the plateau also become much more well-defined at higher temperature. Improvement in the crystallinity is clear from the comparison of the FFTs patterns shown in Figure 2(d), (h), (l) and (p). FFT patterns corresponding to the sample annealed at ~500 °C remains the same as observed from as-sputtered sample [Figure 1(j)]. Ordered (7×7) spots start reappearing for the sample annealed at ~600 °C and the ordering gradually improves with increase in annealing temperatures indicating improvement in crystallinity of the surface. The FFT pattern in Figure 2 (p) (bottom right) shows distinct sharp intense peaks, indicating reappearance of almost a perfect larger area Si(111)-(7×7) surface formed due to recrystallization. There is no evidence of the presence of (2×2) reconstructed chain like structure and (5×5) reconstructed surface structure on the surface. This indicates these metastable structures has been converted into the most stable Si(111)-(7×7) region. T. Hoshino, et. al., observed the gradual growth of (7×7) reconstruction at the boundary of (7×7) and disordered (1×1) regions in Si(111) surface at 600°C [59]. Recently, Shen et.al., using molecular dynamics machine learning simulations found that there are two possible pathways for the formation of the (7×7) structure – first path arises from the growth of a faulted half domain from the metastable (5×5) phase to the final (7×7) structure, while the second path involves the direct formation of the (7×7) reconstruction [50]. This is consistent with our experimental observation that annealing at 600°C transforms the Ar$^+$ ion sputtered unreconstructed surface into the (7×7) reconstructed structure predominantly *via* metastable (2×2), (5×5) which disappears following an annealing at 650°C.

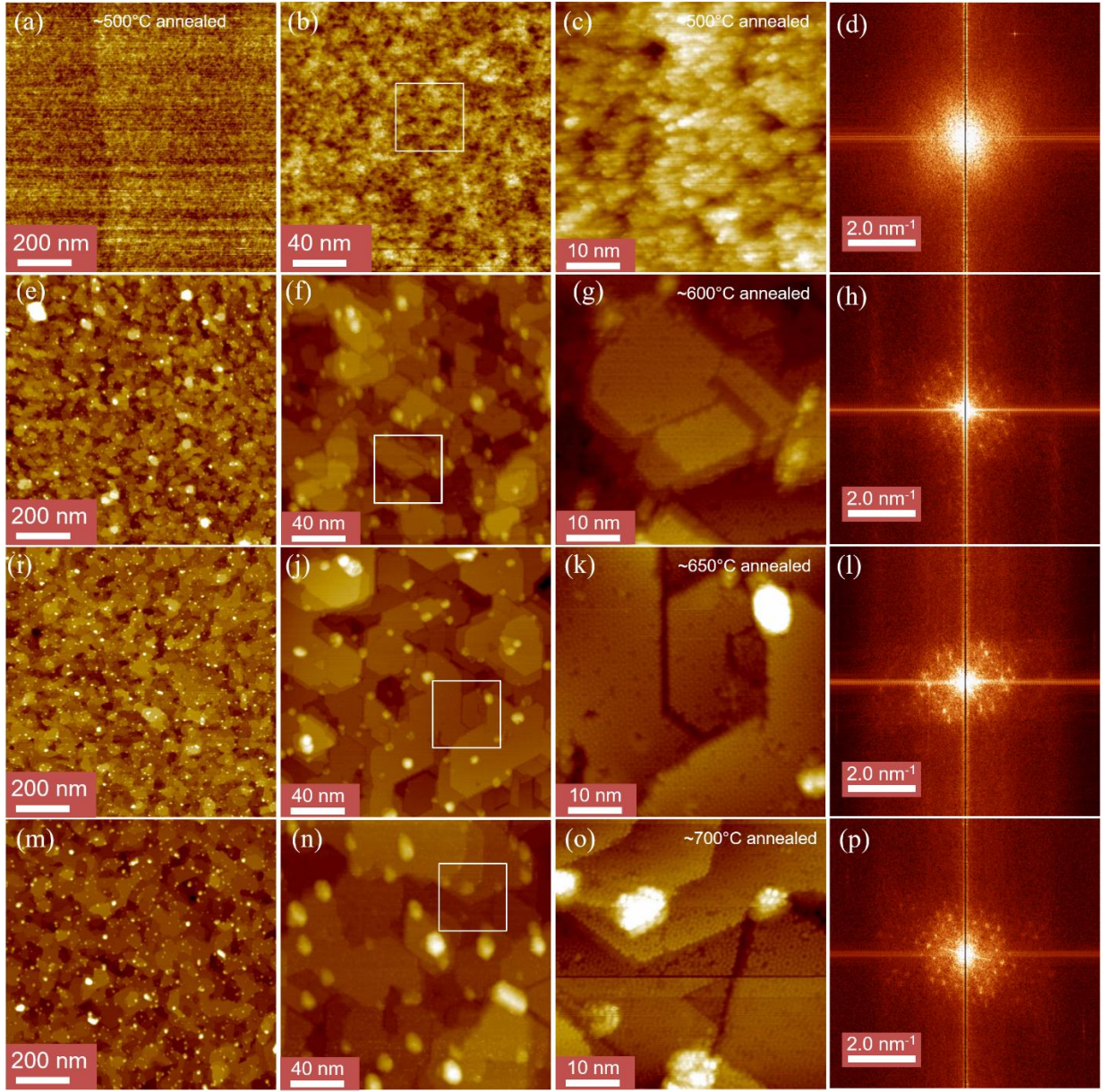

**Figure 2**: STM studies on sputtered-Si(111)-(7×7) surfaces following annealing under UHV show improvement in crystallinity of amorphous surface with increasing annealing temperatures: (a, b, c) ~500 °C, (e, f, g) ~600 °C, (i, j, k) ~650 °C and (m, n, o) ~700 °C of scan areas 1000×1000 nm², 200×200nm² and 50×50 nm², respectively. (d, h, l, p) show the FFT patterns of the 50×50 nm² STM images shown in (c, g, k, o) respectively.

*3.3. Roughness exponent, correlation length of sputtered and post-annealed Si(111) surfaces*

The average rms surface roughness ($\sigma_{avg}$) of the sputtered Si(111)-(7×7) surface following annealing at different temperatures (500–700 °C) have been calculated from the STM images of the different sizes starting from 30×30 nm$^2$ to 2000×2000 nm$^2$ at different locations on the sample surface using WSxM software [60].

The rms static surface roughness $\sigma$ is defined as [61],

$$\sigma = <[h(x,y) - h]^2>^{1/2} \ldots\ldots\ldots \text{Eqn. (1)}$$

where $h(x,y)$ is the surface height at a point $(x,y)$ on the surface, $h$ is the average height and $<>$ denotes the spatial average over positions $(x,y)$ on the surface. The surface is termed as the self-affine fractal surface, if the rms roughness changes with the horizontal sampling length $L$ as [61-65],

$$\sigma \propto L^{\alpha} \ldots\ldots\ldots \text{Eqn. (2)}$$

with the roughness exponent $\alpha$ lying in between 0 and 1. The log-log plots of $\sigma_{avg}$ versus scan size ($L$) for the samples sputtered followed by annealing at different temperatures are shown in Figure 3.

Following annealing at different temperatures, $\sigma_{avg}$ changes in different fashion. While at smaller length scale (below 150 nm), $\sigma_{avg}$ increases sharply and shows a linear variation, it approaches a saturation value ($\sigma_{sat}$) of about 0.6 nm at higher length scales. This behavior is almost the same for all the sputtered samples annealed at temperatures 600 °C – 700 °C. Although, not as prominent as for the samples annealed $\geq$ 600 °C, the sample annealed at 500 °C shows a linear variation in $\sigma_{avg}$ at lower length scales indicating self-affine nature of the surface (with the $0 < \alpha < 1$).

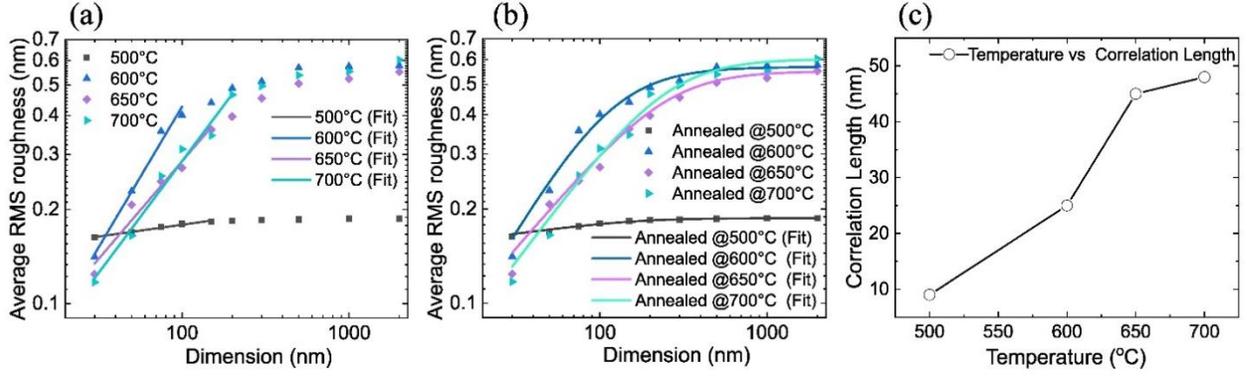

**Figure 3**: Determined rms surface roughness of the Ar$^+$ ion sputtered Si(111)-(7×7) surface annealed at 500 °C – 700 °C plotted as a function of the scan length of the area of interest are shown in Figure 3 (a). Surface roughness exponent values, in each case, have been extracted from the linear fit to the experimental data using Eqn. (2). Figure 3 (b) shows the fit to the experimental data using Eqn. (3). Variation of correlation length with post-annealing temperatures as extracted from the fits is shown in Figure 3 (c).

Roughness exponent ($\alpha$) values as calculated from the slope of the linear variation in $log\sigma_{avg}$ versus $logL$ plots are shown in Table 1. As observed, change in $\alpha$ values do not follow any clear trend with increasing annealing temperatures. This is probably due to random nucleation of silicon nanodots or nanoislands on the surface at higher temperature (as also observed in STM investigation in Figure 2).

| Annealing Temperature (°C) | Roughness exponent (α) | Correlation length (ξ) in nm | Saturation roughness ($\sigma_0$) in nm |
|---|---|---|---|
| 500 | 0.07 ± 0.01 | 9 ± 0.4 | 0.19 |
| 600 | 0.90 ± 0.07 | 25 ± 1 | 0.57 |
| 650 | 0.62 ± 0.07 | 45 ± 2 | 0.55 |
| 700 | 0.71 ± 0.07 | 48 ± 3 | 0.60 |

**Table 1**: The roughness exponent, correlation length and saturation roughness have been extracted from fitting $\sigma_{avg}$ $vs$ $L$ plots of sputtered samples annealed at different temperatures.

For further analysis, the in-plane correlation length (ξ) values are calculated and compared with the observed surface morphologies. ξ measures the average distance between the features in the surface profiles within which the surface variations are correlated. It is related to the surface roughness, which measures the correlation along the direction of surface growth. For $L < \xi$, $\sigma \propto L^{\alpha}$ and for $L > \xi$, $\sigma(L) \rightarrow \sigma_{sat}$. When the correlation length is much larger than the lattice constant ($\xi \gg a$), rms roughness can be expressed as [66, 67],

$$\sigma^2(L) = \frac{\sigma_{sat}^2}{\left(1 + \frac{4\pi\xi^2}{2\alpha L^2}\right)^{\alpha}}$$ …………..Eqn. (3)

Fit to the rms roughness values using Eqn. (3) are shown in Figure 3 (c). Correlation length values, as extracted from corresponding fits, are shown in Table 1. Ar$^+$ ions destroy the crystal structure of Si(111)-(7×7) surface which gradually improves upon annealing, as seen from increasing ξ with increasing annealing temperatures. This is also evident from the STM studies in Figure 2. Single crystalline Si(111)-(7×7) surface transforms into an amorphous surface due to the sputtering process. Post-sputtering annealing process helps to reduce defects and to recrystallize the surface that gradually improves with increasing temperature. This is also confirmed from the power spectral density (PSD) analysis shown in Figure S5 in the supplementary material.

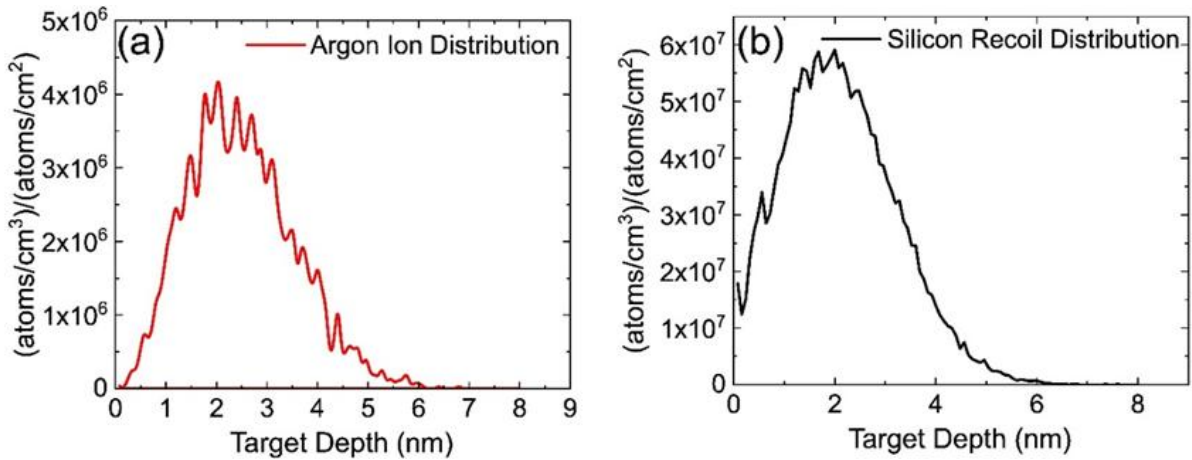

**Figure 4**: Results of the SRIM simulation: (a) depth profile of 0.5 keV Ar$^+$ ions implanted into Si substrate and (b) corresponding recoil distribution of Si atoms.

*3.4. Depth profile and recoil distribution using SRIM simulation*

Figure 4 shows the depth profile of $Ar^+$ ions and corresponding atomic displacements due to collisions of energetic ions with atoms in the sample calculated with the Stopping and Range of Ions in Matter (SRIM) simulations [68, 69]. $Ar^+$ ions of energy 0.5 keV penetrate into the sample and cause atomic displacements of Si(111) sample. Both the distributions show a mean range of about 2.3 nm and about 2 nm, respectively [corresponding to Figure 4(a) and (b)].

The impact of ions, in our case, involves more collisions and sputtering below the top surface layer. As a result along with the (7×7) surface structures, crystalline configuration of Si(111) beneath the surface layer is disorganized. However, as the 0.5 keV $Ar^+$ ions disturb atomic arrangement of silicon atoms only up to ~2 nm from the surface, annealing under UHV can provide enough energy to recrystallize it. For the annealing temperatures ≥ 600 °C, the surface atoms start rearranging and predominantly form reconstructed (7×7) surface (with (2×2) reconstructed chain like structure and (5×5) reconstructed surface structure at some locations). Compared to the surface annealed at 500 °C, the saturation roughness ($\sigma_{sat}$) increases about three times during the recrystallization process. $\sigma_{sat}$ remains almost the same for the surfaces annealed at ≥ 600 °C. This also confirms that the recrystallization process is initiated at temperature ≥ 600 °C. Improvement in the crystal structure is also evident from the increasing correlation length indicating that the surface features can now be correlated for a larger length scale. The average width of these reconstructed smaller flat regions increases (correspondingly the number density decreases) with increasing annealing temperature. As observed from Table 1, the correlation length of the sample annealed at 700°C, remains almost the same as that for the sample annealed at 650°C. For the present experimental conditions, 650 °C is the optimal annealing temperature for recrystallization. Post-annealing under UHV, thus, can be a useful technique to heal undesirable defects created below the surface layer due to sputtering.

## 4. Conclusion

In conclusion, we have studied $Ar^+$ ion sputtered Si(111)-(7×7) surface using *in situ* STM under UHV. An amorphous surface produced *via* sputtering gradually improves in crystallinity with increasing annealing temperatures. Corresponding $\sigma_{avg}$ values estimated from STM images of different scan areas are compared in each case. As observed from high resolution STM studies, sample annealed at ~600 °C under UHV starts showing recrystallization of the amorphous phase and reappearance of (7×7) surface. For the annealing temperatures $\geq$ 600 °C, $log\sigma_{avg}$ varies linearly with $logL$ at lower length scales indicating a self-affine fractal nature of the surface. In-plane correlation length corresponding to each post-annealing temperature also confirms an improvement in crystalline nature. For the present experimental conditions, 650 °C is the optimal annealing temperature for recrystallization.

## Acknowledgments

JCM and DD thank CSIR, India for the fellowships. The experiment was carried out in BND's laboratory when JCM, AR, RB, DD, and BND were all at the Indian Association for the Cultivation of Science, Kolkata, India.